\documentclass[12pt]{iopart}

\usepackage{iopams} 
\usepackage{bm,graphics,graphicx,epsfig,color}

  \newcommand{\apjl}{Astrophys. J. Lett.}

  \newcommand{\mnras}{Mon. Not. R. Astron. Soc.}
   \newcommand{\prd}{Phys. Rev. D}

     \newcommand{\prl}{Phys. Rev. Lett.}

   \newcommand{\aj}{Astron. J.}

 \newcommand{\cqg}{Class. Quantum Grav.}
 \newcommand{\lrr}{Living Rev. Relativ.}

\begin{document}

\title[Graviton mass bounds]{Solar system vs. gravitational-wave bounds on the graviton mass}

\author{Clifford M. Will$^{1,2}$}

\ead{cmw@phys.ufl.edu}
\address{
$^1$ Department of Physics,
University of Florida, Gainesville FL 32611, USA
 \\
$^2$ GReCO, Institut d'Astrophysique de Paris, UMR 7095-CNRS,
Universit\'e Pierre et Marie Curie, 98$^{bis}$ Bd. Arago, 75014 Paris, France
}

\begin{abstract}
The detection of gravitational waves from merging binary black holes has led to a bound on the mass of a hypothetical massive carrier of the gravitational interaction predicted by some modified gravity theories (a massive graviton, for short), corresponding to a bound on the Compton wavelength $\lambda_g > 1.6 \times 10^{13}$ km.  This bound is six times more stringent than a 1988 bound inferred from solar-system dynamics.   Using 30 years of improvements in solar system data, chiefly from missions involving orbiters and probes of planets from Mercury to Saturn, we revisit this bound.  We show that data on the perihelion advance of Mars obtained from the Mars Reconnaissance Orbiter leads to a credible lower bound on $\lambda_g$ between $1.2$ and $2.2 \times 10^{14}$ km, surpassing the gravitational-wave bound by an order of magnitude.   We discuss ways in which each of these competing bounds may improve in the future.
\end{abstract}


\maketitle

\section{Introduction}
\label{sec:intro}

The recent detections of gravitational waves from inspiralling binary black hole systems have made it possible to place bounds on the mass of the ``graviton''. If gravity is propagated by a massive field, then gravitational waves of long wavelength will propagate more slowly than those of short wavelength, leading to a distortion of the gravitational waveform from a binary inspiral that can be bounded using matched filtering \cite{1998PhRvD..57.2061W}.   Data from the discovery source GW150914 resulted in a lower bound $\lambda_g > 10^{13} \, {\rm km}$, where
$\lambda_g$ is the graviton Compton wavelength, corresponding to a mass $m_g <  1.2 \times 10^{-22} \, {\rm eV}/c^2$ \cite{2016PhRvL.116f1102A,2016PhRvL.116v1101A}.   This was improved to $\lambda_g > 1.6 \times 10^{13} \, {\rm km}$ by combining data from the three events GW150914, GW151226 and GW170104  
\cite{2017PhRvL.118v1101A}.

At the time, these bounds improved upon the weaker bound $\lambda_g > 2.8 \times 10^{12} \, {\rm km}$, inferred from solar-system dynamics in a 1988 paper by Talmadge et al.\  \cite{1988PhRvL..61.1159T}.  If the graviton is massive, then in the simplest model, the Newtonian gravitational potential is given by the Yukawa form $(Gm/r) e^{-r/\lambda_g}$ rather than $Gm/r$, where $G$ is Newton's constant.  Such a potential leads to a modification of the relation between orbital period and semimajor axis and to additional contributions to perihelion advances of the planets.  The limit inferred from  \cite{1988PhRvL..61.1159T} derived mostly from orbital period data  \cite{1998PhRvD..57.2061W}.

Much has changed since 1988.   The precision of our knowledge of planetary motion has steadily improved as a result of high-precision radar tracking of planets and spacecraft, improvements in the measurement of planetary and asteroid masses, increasingly precise measurements of the Earth-Moon orbit using lunar laser ranging and development of improved ephemeris computer codes.  In particular, major advances in precision have come from the recent array of planetary orbiters tracked precisely via range or Doppler radar  and via Very Long Baseline Interferometry.  These include, Mars Express (2003 - ), Venus Express (2006 - 2014), Mars Reconnaissance Orbiter (MRO) (2006 - ), the Cassini orbiter of Saturn (2004 - 2017) and Mercury MESSENGER orbiter (2011 - 2015).   

In particular, the perihelion advance of Mercury and Mars are now known to a few parts in $10^5$, and that of Saturn is known to a few percent.    To leading order in the ratio $(a/\lambda_g)$, where $a$ is the orbital semimajor axis, the perihelion advance per orbit induced by a massive graviton is given by
\begin{equation}
\Delta \varpi =   \pi \left ( \frac{a}{\lambda_g} \right )^2 (1-e^2)^{-1/2} \,, 
\label{eq:perihelion}
\end{equation}
where $\varpi$ is the longitude of perihelion measured from a fixed reference direction, and $e$ is the orbital eccentricity.  Recall that the advance induced by general relativity (GR) is $\Delta \varpi_{\rm GR} = 6 \pi Gm/c^2a(1-e^2)$, where $m$ is the mass of the Sun, and $G$ and $c$ are the Newtonian gravitational constant and speed of light, respectively.  While the GR precession is largest for orbits close  to the Sun, the massive graviton effect grows with distance from the Sun.

In this paper we survey the latest analyses of solar system data, and show that the best bound on $\lambda_g$ comes from the perihelion advance of Mars incorporating MRO data.  It is given by the range 
\begin{equation}
\lambda_g > (1.2 - 2.2) \times 10^{14} \, {\rm km} \,, \quad m_g < (6 - 10) \times 10^{-24} \, {\rm eV/c^2} \,,
\end{equation}
depending on the specific analysis.   This bound is almost two orders of magnitude larger than that inferred in 1988 and an order of magnitude better than the LIGO bound.   It is significantly stronger than bounds inferred from binary-pulsar data ($\lambda_g > 10^{10} \, {\rm km}$) \cite{2002PhRvD..65d4022F}, and stronger by almost an order of magnitude than model-dependent bounds inferred from the stability of the Schwarzschild and Kerr metrics  \cite{2013PhRvD..88b3514B}.   Various studies have shown that a stronger bound of $\lambda_g > 10^{16} \, {\rm km}$ could be obtained from gravitational-wave detections by the LISA space antenna \cite{1998PhRvD..57.2061W,2005PhRvD..71h4025B,2009PhRvD..80d4002S,2009CQGra..26o5002A}.  
A frequently quoted bound is  $\lambda_g > (6 - 9) \times 10^{19} \, {\rm km}$ from galactic and
cluster dynamics
\cite{1965PThPS..65E.261H,1973CaJPh..51..431H,1974PhRvD...9.1119G,2018PhLB..778..325D};  
however, in view of the uncertainties
related to the amount of dark matter in the universe, and the fact that massive gravity theories frequently include other modifications of gravity on large scales, this latter
bound should be viewed as model-dependent.

The rest of this paper provides details.  Section \ref{sec:effects} briefly summarizes the derivations of the leading effects of a massive graviton in the solar system.  In Sec.\ \ref{sec:bounds}, we obtain bounds on the graviton mass from a range of analyses of solar system ephemeris data.   Section \ref{sec:discussion} discusses prospects for future improvements.  We remind the reader that the term graviton is merely shorthand for a massive gravitational field; quantum gravity plays no role in this discussion.

\section{Effects of a massive graviton in the solar system}
\label{sec:effects}

In the Newtonian limit of a theory with a massive graviton, the gravitational potential in a system of $N$ bodies is given, in the simplest model, by
\begin{eqnarray}
U &=& \sum_a \frac{Gm_a}{|\bm{x} - \bm{x}_a|} e^{-|\bm{x} - \bm{x}_a|/\lambda_g}  
\nonumber \\
& =& \sum_a \frac{Gm_a}{|\bm{x} - \bm{x}_a|}  + \frac{1}{2} \sum_a \frac{Gm_a}{\lambda_g^2}  |\bm{x} - \bm{x}_a| + O(Gmr^2/\lambda_g^3)\,,
\label{eq:eom0}
\end{eqnarray}
where we have dropped an irrelevant constant in the second line.  
 Since we already know that $\lambda_g > 10^4$ astronomical units (au), we have expanded the potential in powers of $\lambda_g^{-1}$, retaining only the leading non-trivial correction term.  For the two-body problem, the equation of motion is given by
\begin{equation}
\frac{d\bm{v}}{dt} = - \frac{Gm {\bm n}}{r^2} \left ( 1- \frac{1}{2} \frac{r^2}{\lambda_g^2} \right ) \,,
\label{eq:eom}
\end{equation}
where $m$ is the sum of the two masses.  Carrying out the standard orbital perturbation methods as outlined in Section 3.3.2  of  \cite{2014grav.book.....P}, it is a simple exercise to show that the pericenter advance is given by Eq.\ (\ref{eq:perihelion}).  For a nearly circular orbit, Eq.\ (\ref{eq:eom}) leads to an orbital angular velocity given by $\omega^2 = (Gm/r^3)[1 - r^2/2\lambda_g^2]$, and thus to an orbital period given by 
\begin{equation}
P \approx 2\pi \left (\frac{a^3}{Gm} \right )^{1/2} \left ( 1+ \frac{1}{4}\frac{a^2}{\lambda_g^2}  \right ) \,,
\label{eq:period}
\end{equation}
where $a$ is the semimajor axis.
One can also evaluate the impact of the Yukawa modification on the perturbation of a given planet by another planet.  Following the same method used in \cite{2014grav.book.....P} to determine the Newtonian effect of a third body, we find that $\Delta \varpi = (3\pi/4)(a/\lambda_g)^2 (m_3/m)(a/a_3)(1-e^2)^{1/2}$, where $m_3$ and $a_3$ are the mass and semimajor axis of the third body, assumed for simplicity to be in a circular orbit.  For any pairs of planets this effect is several orders of magnitude smaller than the leading effect displayed in Eq.\ (\ref{eq:perihelion}), and thus will play no role in bounding $\lambda_g$. 

If the residual uncertainty in the perihelion advance of a given planet after planetary perturbations and standard GR effects have been modelled is given by $\sigma(\dot{\varpi})$, then the lower bound on $\lambda_g$ can be expressed in the form
\begin{equation}
\lambda_g >  \left \{
\begin{array}{l}
3.82 \times 10^{12} \, {\rm km} \, \left [\frac{\displaystyle \tilde{a}^{1/4}}{\displaystyle \sigma(\dot{\varpi})^{1/2} (1-e^2)^{1/4}} \right ]\,, 
 \\
 \\
6.17 \times 10^{11}\, {\rm km} \, \left [\frac{\displaystyle \tilde{a}^{3/2}(1-e^2)^{1/4}}{\displaystyle \sigma_{\rm rel}(\dot{\varpi})^{1/2} } \right ] \,, 
\end{array} 
\right .
\label{eq:bound1}
\end{equation}
where $\tilde{a}$ is the semimajor axis of the planet in astronomical units, $\sigma(\dot{\varpi})$ is the uncertainty in $\dot{\varpi}$ in milliarcseconds per year, and $\sigma_{\rm rel}(\dot{\varpi})$ is the same uncertainty divided by the GR advance rate for that planet.  In the latter case, this could be the uncertainty in the PPN parameter coefficient $(2 + 2\gamma -\beta)/3$ inferred from that planet's perihelion advance.

The violation of Kepler's third law exhibited in Eq.\ (\ref{eq:period}) can also be used to bound $\lambda_g$.  We define the parameter $\eta$ by \cite{1988PhRvL..61.1159T,1998PhRvD..57.2061W}
\begin{equation}
\eta \equiv  \left ( \frac{n^2 a^3}{n_\oplus^2 a_\oplus^3} \right )^{1/3} -1 \,,
\label{eq:eta}
\end{equation}
where $n = 2\pi/P$ is the measured ``mean motion'' and $a$ is the measured semimajor axis of a given planetary orbit, with the subscript $\oplus$ denoting the Earth.   Then a bound on $\lambda_g$ can be expressed in the form
\begin{equation}
\lambda_g > 1.5 \times 10^8 \, {\rm km} \, \left ( \frac{1-\tilde{a}^2 }{6\eta} \right )^{1/2}\,.
\end{equation}

\section{Bounds on $\lambda_g$ from solar-system data}
\label{sec:bounds}

The tightest bounds on $\lambda_g$ come from data on the perihelion advances of the planets, notably Earth, Mars and Saturn.  Two analyses 
\cite{2013MNRAS.432.3431P,2016arXiv160100947F} tabulated the residual uncertainties in the perihelion advances of Mercury through Saturn, after the effects of planetary and asteroid perturbations and standard GR were taken into account.  Table \ref{tab:bounds1} lists the uncertainties in $\sigma(\dot{\varpi})$ and the bounds on $\lambda_g$ inferred from the first of Eqs.\ (\ref{eq:bound1}).
Note that the strongest bounds come from Mars and Saturn, where the MRO and Cassini missions, respectively, have led to improved orbital knowledge.

\begin{table}[t] 
\centering
\caption{Bounds on $\lambda_g$} 
\vskip 12pt
\begin{tabular}{@{}lcc@{}} 
\hline 
Planet&$\sigma(\dot{\varpi})$ (mas/yr)&$\lambda_g$ bound ($10^{14}$ km)\\ 
\noalign{\smallskip}
\hline 
\multicolumn{3}{l}{{\em Data from Table 4 of \cite{2013MNRAS.432.3431P}}}\\
\noalign{\smallskip}
Mercury&$0.03$&$0.18$\\
\noalign{\smallskip}
Venus&$0.016$&$0.28$\\
\noalign{\smallskip}
Earth&$0.0019$&$0.88$\\
\noalign{\smallskip}
Mars&$0.00037$&$2.21$\\
\noalign{\smallskip}
Jupiter&$0.28$&$0.11$\\ 
\noalign{\smallskip}
Saturn&$0.0047$&$0.98$\\ 
\noalign{\smallskip}
\multicolumn{3}{l}{{\em Data from \cite{2016arXiv160100947F}}}\\
Mercury&$0.02$&$0.22$\\
\noalign{\smallskip}
Saturn&$0.026$&$0.42$\\ 
\hline
\end{tabular}
\label{tab:bounds1}
\end{table}

In other analyses, fits to the data using the PPN framework were carried out, quoting bounds on the parameters $\gamma$ and $\beta$.  Most of the bounds were dominated by Mercury data, but one analysis focussed on bounds that could be inferred from MRO data on Mars.  In each case, we estimated the uncertainty in $\sigma_{\rm rel}(\dot{\varpi})$ by simply adding the uncertainties in $\gamma$ and $\beta$, weighted by the numerical coefficient in the PPN formula for the perihelion advance, namely
\begin{equation}
\sigma_{\rm rel}(\dot{\varpi}) \equiv \frac{2}{3} \sigma(\gamma)+ \frac{1}{3} \sigma(\beta) \,.
\end{equation}
The results for $\sigma_{\rm rel}(\dot{\varpi})$ and the bounds inferred for $\lambda_g$ are shown in Table \ref{tab:bounds2}.  

The strongest bounds, between $1.2$ and $2.2 \times 10^{14}$ km, come from Mars, exploiting its distance from the Sun and the measurement accuracy derived from MRO.  Close behind are bounds from Earth and Saturn.   Mercury contributes a substantially weaker bound.

\begin{table}[t] 
\centering
\caption{Bounds on $\lambda_g$. Asterisk denotes analyses where $\gamma$ was constrained by measurements of the Shapiro delay using Cassini.} 
\vskip 12pt
\begin{tabular}{@{}lcccc@{}} 
\hline 
&&&&$\lambda_g$ bound\\
Reference&$\sigma(\gamma) \times 10^5$&$\sigma(\beta) \times 10^5$&$\sigma(\dot{\varpi}_{\rm rel}) \times 10^5$&($10^{14}$ km)\\ 
\noalign{\smallskip}
\hline 
\noalign{\smallskip}
{\em Mercury}&\\
\noalign{\smallskip}
Pitjeva \& Pitjev \cite{2013MNRAS.432.3431P}&6.0&3.0&5.0&0.21 \\
Verma et al. \cite{2014A&A...561A.115V}&2.5&2.5&2.5&0.29\\
Fienga et al. \cite{2015CeMDA.123..325F}&5.0&6.9&5.6&0.20 \\
Park et al. \cite{2017AJ....153..121P}&2.3*&3.9&2.8&0.28\\
Genova et al.\cite{2018NatCo...9..289G}&2.3*&1.8&2.1&0.32 \\
{\em Mars}\\
Konopliv et al. \cite{2011Icar..211..401K}&2.3*&24&9.5&1.18 \\
\hline
\end{tabular}
\label{tab:bounds2}
\end{table}

Talmage et al.\ \cite{1988PhRvL..61.1159T} used solar-system data on orbital periods and semimajor axes to establish bounds on the parameter $\eta$ in Eq.\ (\ref{eq:eta}).  The best bound on $\lambda_g$ came from Mars, with the $2\sigma$ one-sided bound $\eta > -6.5 \times 10^{-10}$ (see Table 1 of  \cite{1988PhRvL..61.1159T} and Table 4 of \cite{1998PhRvD..57.2061W}), leading to $\lambda_g > 2.8 \times 10^{12}$ km. 
We are not aware of more recent analyses of solar-system data that have taken this approach to bounding the massive graviton.  Note that to reach the level of a few times $10^{14}$ km, the bound on $\eta$ for Mars would have to be improved by three orders of magnitude, corresponding to uncertainties in the semimajor axis at the level of centimeters, and in the orbital period at the level of tens of microseconds.
 
Finally, the constant radial acceleration in Eq.\ (\ref{eq:eom}) induced by the massive graviton at leading order in $r/\lambda_g$ is reminiscent of the so-called ``Pioneer anomaly'', an apparent anomalous constant acceleration of $(8.74 \pm 1.33) \times 10^{-10} \, {\rm m/s}^2$ inferred from Doppler tracking data on the Pioneer 10 and 11 spacecraft.   It is now generally accepted that the origin of this acceleration was the anisotropic radiation of thermal energy from the radiothermal generators on board the spacecraft, and not new physics \cite{2010LRR....13....4T,2012PhRvL.108x1101T}.  Nevertheless, if we take the measurement error as the upper limit on any anomalous acceleration $\delta a$ that could be attributed to a massive graviton, then the lower bound on $\lambda_g$ is given by
\begin{equation}
\lambda_g > \left ( \frac{Gm}{2 \delta a} \right )^{1/2} \sim 7 \times 10^{11} \, {\rm km} \,,
\end{equation}
not competitive with other solar-system bounds.    
 
\section{Discussion}
\label{sec:discussion}

We have estimated a bound on the graviton mass using results of analyses of solar-system data that were carried out for other purposes, such as measuring the PPN parameters $\gamma$ and $\beta$.  It would be desirable to carry out such analyses by systematically including in the equations of motion of the ephemeris codes the effects of a massive graviton as displayed in Eq.\ (\ref{eq:eom0})  (see \cite{2008AIPC..977..254S} for a preliminary analysis in the context of a constant Pioneer-type acceleration).  Only then can one assess the effects of correlations among the various parameters on the bound that can be obtained for $\lambda_g$.  Because analyses of solar system data are dominated by systematic effects, it is difficult to assess a priori whether the results will be better or worse than the estimates made in this paper.  

Progress in tightening the bound on $\lambda_g$ in the near future is likely to be slow.   For ground-based gravitational-wave interferometers, the bound scales roughly as \cite{1998PhRvD..57.2061W}
\begin{equation}
\lambda_g \sim S_0^{-1/4}  f_0^{-1/3} {\cal M}^{11/12}\,,
\end{equation}
where $S_0$ and $f_0$ are measures of the noise ``floor'' and the frequency of peak sensitivity of the detector, respectively, and ${\cal M}$ is the ``chirp mass'' of the source.\footnote{This estimate is based on 1998-era analytic models for the advanced LIGO noise curve}  Note that the bound is largely independent of the source distance or signal-to-noise ratio; this is because, while parameter estimation accuracy for a given source decreases linearly with distance (i.e.\ with decreasing signal-to-noise ratio), the cumulative effect of the massive graviton on the propagation of the signal increases linearly with distance.   The bound increases roughly linearly with chirp mass, however success will be limited by the fact that more massive binaries will merge before entering the sensitive band of the ground-based detectors (see also
 \cite{2011PhRvD..83h2002D}).   Because it is of lower mass, the neutron-star merger event GW170817 will not give a competitive bound.  From the time difference of $\Delta t = 1.74$ s between the gravitational and electromagnetic signals and the distance $D =  40$ Mpc \cite{2017ApJ...848L..13A}, one can estimate $\lambda_g > f^{-1} (D/\Delta t)^{1/2} \sim 10^{10}$ km, where $f \sim 100$ Hz is the gravitational-wave frequency.

Similarly, because the bound depends on the square root of measurement uncertainty, solar-system data are unlikely to yield a dramatic improvement in the near future.  The BepiColombo mission to Mercury, scheduled for launch in late 2018, may decrease the uncertainty in the perihelion advance of Mercury by a factor of ten, but as Table 2 indicates, an {\em additional} factor of ten would be needed to bring the bound on $\lambda_g$ within striking distance of the bound obtained from Mars.  Mercury is good for testing GR, but not so good for testing a massive graviton.

 \ack
This work was supported in part by the National Science Foundation,
Grant No.\ PHY 16--00188.      



\section*{References}


\providecommand{\newblock}{}

\end{document}